\documentclass[twocolumn,linenumbers]{aastex631}

\usepackage{amsmath} 
\usepackage{xspace}
\usepackage{cases}
\usepackage[flushleft]{threeparttable}
\usepackage{color}
\usepackage{hyperref} 

\newcommand{\galname}{A2744-z7DLA\xspace}

\newcommand{\galfit}{\texttt{GALFIT}\xspace}
\newcommand{\bagpipes}{\texttt{Bagpipes}\xspace}
\newcommand{\msaexp}{\texttt{msaexp}\xspace}

\newcommand{\lmfit}{\texttt{LMFIT}\xspace}
\newcommand{\cigale}{\texttt{CIGALE}\xspace}
\newcommand{\galfits}{\texttt{GalfitS}\xspace}

\newcommand{\grizli}{\texttt{grizli}\xspace}
\newcommand{\psfex}{\texttt{PSFEx}\xspace}
\newcommand{\sextractor}{\texttt{SExtractor}\xspace}
\newcommand{\I}{\text{\uppercase\expandafter{\romannumeral 1}}}
\newcommand{\II}{\text{\uppercase\expandafter{\romannumeral 2}}}
\newcommand{\III}{\text{\uppercase\expandafter{\romannumeral 3}}}

\newcommand{\narrow}{0ex}

\graphicspath{{./}{figures/}}

\begin{document}
\nolinenumbers
\title{
Discovery of a gas-enshrouded broad-line AGN at $z\sim7$
}

\author[0009-0006-1255-9567]{Qianqiao Zhou}
\affiliation{School of Astronomy and Space Science, University of Chinese Academy of Sciences (UCAS), Beijing 100049, China}

\author[0000-0002-9373-3865]{Xin Wang}
\affiliation{School of Astronomy and Space Science, University of Chinese Academy of Sciences (UCAS), Beijing 100049, China}
\affil{National Astronomical Observatories, Chinese Academy of Sciences, Beijing 100101, China}
\affil{Institute for Frontiers in Astronomy and Astrophysics, Beijing Normal University,  Beijing 102206, China}

\author[0009-0004-7133-9375]{Hang Zhou}
\affil{School of Astronomy and Space Science, University of Chinese Academy of Sciences (UCAS), Beijing 100049, China}

\author[0000-0002-3331-9590]{Emanuele Daddi}
\affiliation{Laboratoire AIM, CEA/DSM-CNRS-Universit\'e Paris Diderot, IRFU/Service d'Astrophysique, B\^at. 709, CEA Saclay, F-91191 Gif-sur-Yvette Cedex, France}

\author[0000-0001-6947-5846]{Luis C. Ho}
\affil{Kavli Institute for Astronomy and Astrophysics, Peking University, Beijing 100871, China}
\affil{Department of Astronomy, School of Physics, Peking University, Beijing 100871, China}

\author[0009-0007-6655-366X]{Shengzhe Wang}
\affiliation{School of Astronomy and Space Science, University of Chinese Academy of Sciences (UCAS), Beijing 100049, China}
\affil{National Astronomical Observatories, Chinese Academy of Sciences, Beijing 100101, China}

\author[0000-0001-8496-4162]{Ruancun Li}
\affiliation{Max-Planck-Institut f{\"u}r extraterrestrische Physik, Gie{\ss}enbachstra{\ss}e 1, 85748 Garching bei M{\"u}nchen, Germany}

\author[0000-0002-2178-5471]{Zuyi Chen}
\affiliation{Cosmic Dawn Center (DAWN)}
\affiliation{Niels Bohr Institute, University of Copenhagen, Jagtvej 128, 2200 Copenhagen N, Denmark}

\author[0000-0003-0202-0534]{Cheng Cheng}
\affiliation{Chinese Academy of Sciences South America Center for Astronomy, National Astronomical Observatories, CAS, Beijing 100101, China}

\author[0000-0002-1660-9502]{Xihan Ji}
\affiliation{Kavli Institute for Cosmology, University of Cambridge, Madingley Road, Cambridge CB3 0HA, UK}

\author[0009-0005-3823-9302]{Yuxuan Pang}
\affiliation{School of Astronomy and Space Science, University of Chinese Academy of Sciences (UCAS), Beijing 100049, China}

\author[0000-0002-5815-2387]{Mengting Ju}
\affiliation{School of Astronomy and Space Science, University of Chinese Academy of Sciences (UCAS), Beijing 100049, China}

\correspondingauthor{Xin Wang}
\email{xwang@ucas.ac.cn}

\begin{abstract}
\nolinenumbers
The Lyman-alpha (Ly$\alpha$) absorption profile in star-forming galaxies serves as a powerful tracer of the extended, dense neutral hydrogen in their surroundings during the Epoch of Reionization (EoR).
We report a unique galaxy, \galname, at $z\approx 6.87$ gravitationally lensed by the foreground galaxy cluster Abell 2744, which exhibits both moderate Ly$\alpha$ emission and damped Ly$\alpha$ absorption, suggesting the presence of a dense neutral hydrogen environment. Our analysis suggests that 
the UV continuum turnover near Ly$\alpha$ is more likely shaped by a damped Ly$\alpha$ system rather than nebular continuum from two photon process.
We analyze the physical properties of \galname with spectroscopic and photometric data from the JWST and the HST.  The galaxy shows a compact morphology ($r_e \sim 0.3\ {\rm kpc}$) and a broadened H$\alpha$ emission line, suggesting possible AGN activity. The broad component of H$\alpha$ has a FWHM of $2721 \pm 200\ {\rm km\ s^{-1}}$, corresponding to a black hole mass of $M_{\mathrm{BH}}=2.90^{+2.35}_{-1.28}\times 10^7 M_\odot$ and a black hole–to–stellar mass ratio of $\log (M_{\rm BH}/M_{\rm stellar}) = -1.58^{+0.45}_{-0.34}$. The Balmer decrement ($\rm H\alpha/H\beta$) yields a dust attenuation of $\rm A_V \approx 1.15 \pm 0.23$, indicating that this system is less dust-rich than some “little red dots”.
Furthermore, we perform SED fitting using both stellar and AGN models. The results show that the UV and optical wavelengths are dominated by star-forming regions, while the AGN component contributes primarily at longer wavelengths. This work provides new insights into the interplay between star formation, neutral gas, and potential AGN activity in galaxies during the EoR.

\end{abstract}


\section{Introduction} \label{sect:intro}

The distribution and physical state of the neutral gas surrounding high redshift galaxies provide key insight into their star forming activities and evolutionary processes \citep{reviewDLA}. A direct tracer of neutral hydrogen is the 21\,cm hyperfine transition; however, the weakness of H \I\ 21cm emission at cosmological distances makes it extremely difficult to detect from individual galaxies \citep{back1,back2}.

An alternative indicator of neutral hydrogen during the Epoch of Reionization (EoR) is the absorption feature of Lyman-alpha (Ly$\alpha$) emission caused by extended, dense gas surrounding star-forming galaxies \citep{strong_damped}. Among Ly$\alpha$ absorbers, damped Ly$\alpha$ absorption (DLA) systems, with HI column densities $\rm N_{HI} > 2 \times 10^{20}  \text{cm}^{-2}$, are fundamentally different because the hydrogen in these systems is predominantly neutral \citep{reviewDLA}. The absorption strength is determined by the density and spatial distribution of neutral hydrogen along the line of sight, making it a useful tool for constraining the characteristics of the high-redshift intergalactic medium (IGM) \citep{JWST_damping}. Studying Ly$\alpha$ absorbers helps to elucidate the relationship between their host galaxies and the surrounding neutral gas \citep{lya_intro1,lya_intro2}. Moreover, DLA systems provide valuable insights into the chemical evolution and the interplay between neutral gas and star-forming activity in high-redshift galaxies \citep{reviewDLA}.

The launch of the James Webb Space Telescope (JWST) has enabled us to explore the DLA during the EoR. So far, JWST/NIRSpec, especially prism spectra, has revealed numerous DLA systems at $z > 5$, which are located in almost fully neutral IGM \citep{DLA1, DLA2, DLA3, DLA4,DLA5,DLA6}. However, due to the low spectral resolution of the prism around $1\ \mathrm{\mu m}$, the DLA feature may be degenerate with other features that alter the shape of the UV continuum \citep{JWST_damping}. For example, the UV continuum is sampled by only a few pixels, allowing moderate Ly$\alpha$ emission to contaminate the continuum \citep{moderateLya1, moderateLya2, moderateLya3, moderateLya4}. Further observations, combined with models that incorporate more realistic distributions of neutral hydrogen gas in ionized bubbles and IGM, are required to robustly constrain the cosmic reionization timeline.

Thanks to the exceptional sensitivity of JWST, a very special kind of galaxy has been discovered—the ``little red dots" (LRDs). These are optically red sources with very compact sizes, and their spectra often exhibit broad permitted emission lines (FWHM $>1500\ \mathrm{km\ s^{-1}}$) \citep{LRD_discover1,LRD_discover2,LRDcandidate,LRD_discover4}. Nevertheless, a sample of LRDs with narrow emission lines and without broad H$\alpha$ lines indicates that the Little Red Dot population may comprise a heterogeneous mixture of sources with different physical natures \citep{narrow-line}. Most of the LRDs identified by JWST are located at redshifts $z \sim 4$–$8$ or even higher \citep{z10discovery}, but the statistical search for LRDs is being extended to $z<4$ in wide fields \citep{euclid, lin_model,ma2025counting,lord}. LRDs exhibit some puzzling properties, such as a significant Balmer break or a V-shaped spectral energy distribution (SED) \citep[e.g.,][]{Balmer_break1,LRD_discover2,balmer_break2,LRDcandidate,LRD_discover5, setton2024,hviding2025}, absorption features superimposed on broad H$\alpha$ and/or H$\beta$ emission lines \citep[e.g.,][]{Balmer_absorption1, Balmer_absorption2, LRD_discover4,blackthunder2,blackthunder_jades}. Based on multi-epoch observations \citep{kokubo2024challenging}, most LRDs exhibit little variability and X-ray weakness \citep{ananna2024x,yue2024stacking,lambrides2024case,maiolino2025jwst}, possibly as a result of super-Eddington accretion onto black holes \citep{Tentative_Detection, variability,BlackTHUNDER}. 

Thorough analysis of low-redshift analogs of high-redshift LRDs implies an atypical stellar and AGN model \citep[e.g.,][]{lord,lin_model,local}. Previous studies aimed at explaining the spectroscopic features of LRDs emphasize that the extremely dense gas may cause the V-shaped SED \citep{dense_gas, black_hole_star, BlackTHUNDER, over_mass1,de2025remarkable,taylor2025capers}. \cite{dense_gas} explored how dense gas and AGN activity could account for the Balmer break and absorption features, ensuring that the inferred stellar mass remains consistent with structure formation models \citep{over1,over2}. However, \citet{liu2025balmer} argue that a super-Eddington accretion system can naturally give rise to both the Balmer break and the red optical color, eliminating the need for external gas absorption or dust. 

It has occurred to us that the possible physical conditions of LRDs may help explain certain high-redshift DLA systems, as they share some similar features. For instance, GS9422 in the Hubble Ultra Deep Field \citep[HUDF;][]{GS9422_field} is a low-mass, compact galaxy. The JWST Advanced Deep Extragalactic Survey \citep[JADES][PIs Rieke and Lützgendorf]{GS9422_JADES} has determined its spectroscopic redshift to be 5.943. Several studies have discussed the nature of the UV continuum and Ly$\alpha$ emission of GS9422 \citep{GS9422_1, GS9422_2, GS9422_3, GS9422_6, nebular_cont,nebular_cont2, nebular_cont3}. \citet{GS9422_2,GS9422_3} agree that an AGN is required to reproduce the UV continuum of GS9422, although \cite{nebular_cont,nebular_cont2,nebular_cont3} suggests that nebular emission powered by stellar populations with a top-heavy IMF is also a plausible source. In this work, we present a similar galaxy, \galname, which exhibits both a curved Lyman break and moderate Ly$\alpha$ emission. In addition, \galname shows broad H$\alpha$ emission and strong rest-optical lines, along with a compact morphology and blue rest-optical colors. These characteristics make it a compelling system to link with both DLA galaxies and LRDs.

This paper is organized as follows. In Section \ref{sect:data}, we provide a brief overview of the observation and data reduction process. Section \ref{sect:rslt} presents the basic results and analysis of \galname, including its physical properties derived from SED fitting, UV continuum and Ly$\alpha$ features, as well as emission diagnostics. In Section \ref{sec:discuss}, we discuss the results in detail. Finally, the conclusions and future prospects are summarized in Section \ref{sec:conclu}. Throughout this paper, we adopt a $\Lambda$CDM cosmology with $\mathrm{\Omega_{m}} = 0.3$, $\Omega_{\Lambda} = 0.7$, and $\mathrm{H_0} = 70\ \mathrm{km\ s^{-1}\ Mpc^{-1}}$. The cosmic distances used in this study are calculated using the Cosmology Calculator \citep{cosmic_cal}.
\begin{table}
\begin{threeparttable}
\caption{Physical Properties of Galaxy \galname}
\begin{tabular}{lcccc}
 \hline\hline 
 Parameters  &  Values \\
 \hline \noalign {\smallskip}
R.A. [deg] & 3.580446 \\ [\narrow]
Decl. [deg] & -30.404990 \\ [\narrow]
$z_{\rm spec}$ & 6.87 \\ [\narrow]
$\mu$ (magnification) & $2.26_{-0.14}^{+0.14}$ \\[\narrow]
$r_e$ [kpc] & $0.36\pm0.01$ \\[\narrow]
 \hline \noalign {\smallskip}

\multicolumn{2}{c}{Direct spectral results}   \\ 
$\rm A_{V}$ [mag] & $1.15\pm0.23$ \\ [\narrow]
$\beta$ (UV slope)& $-2.46^{+0.03}_{-0.03}$ \\ [\narrow]
SFR (H$\beta$) [$M_{\odot}$/yr]  & $13.21\pm3.94$ \\ [\narrow]

\hline \noalign {\smallskip}
\multicolumn{2}{c}{SED fitting results}   \\ 
$\log(M_{\ast}/M_{\odot})$ & $8.77^{+0.24}_{-0.11}$ \\ [\narrow]
SFR (SED fitting) $[M_{\odot}/yr]$ & $2.81\pm0.36$ \\ [\narrow]
$\log(Z_{\ast}/Z_{\odot})$ & $-1.01\pm0.02$ \\ [\narrow]
$t_{\rm age}$ [$Myr$] & $12.37_{-1.74}^{+4.21}$ \\ [\narrow]

\hline \noalign {\smallskip}
\multicolumn{2}{c}{Observed emission line ratios}   \\ 
$\rm H\alpha/\rm H\beta$ & $3.90\pm0.25$ \\ [\narrow]
[O \III] $\lambda 5007$/[O \III] $\lambda 4363$ & $26.56\pm9.60$ \\ [\narrow]
[O \III] $\lambda 5007$/[O \II] $\lambda\lambda 3727,3729$ & $>13.22$ \\ [\narrow]
[O \III] $\lambda 5007$/H$\beta$ & $3.11\pm0.24$ \\ [\narrow]
[O \III] $\lambda 4363$/H$\gamma$ & $0.19\pm0.07$ \\ [\narrow]
[Ne \III] $\lambda 3869$/[O \II] $\lambda\lambda 3727,3729$ & $>1.79$ \\ [\narrow]

\hline \noalign {\smallskip}
\multicolumn{2}{c}{DLA analyses}   \\ 
$\log \rm EW_{Ly\alpha}\ [\mathrm{\AA}]$ & $1.89_{-0.07}^{+0.08}$ \\ [\narrow]
$\log \rm N_{H\I}\ [cm^{-2}]$ & $22.20_{-0.37}^{+0.32}$ \\ [\narrow]
$\Delta v\ [10^3\ \mathrm{km\,s^{-1}}]$ & $-14.95_{-3.11}^{+3.19}$ \\ [\narrow]
$f_c$ & $0.24_{-0.05}^{+0.08}$ \\ [\narrow]
$\rm R_{ion}\ [Mpc]$ & $1.17^{+0.56}_{-0.47}$ \\ [\narrow]
 \hline \noalign {\smallskip}
 \end{tabular}

\begin{tablenotes}
      \item[NOTE ---] The effective radius is measured using \galfits, with magnification correction applied.
 The magnification factor $\mu$ is derived from the CATS 4.1 lensing model \citep{lens_model}. The star formation rate from H$\beta$ emission, SFR(H$\beta$), is calculated using the relation $\text{SFR}(\text{H}\beta) = 4.65 \times 10^{-42}\, L_{\text{H}\beta}\,[{\rm erg\ s^{-1}}] \times 2.86\ [M_\odot\ {\rm yr^{-1}}]$ \citep{SFR_Hb, SFR_Hb2}. The stellar mass, $\log(M_{\ast}/M_{\odot})$, SFR (SED fitting), $\log(Z_{\ast}/Z_{\odot})$, and $t_{\rm age}$ [Myr] are derived from \bagpipes\ SED fitting. The observed line ratios are reported prior to dust correction. The parameter $\beta$ denotes the UV continuum slope. All quoted uncertainties correspond to $1\sigma$ confidence intervals.
\end{tablenotes}

\label{tab:gal}
\end{threeparttable}
\end{table}


\section{Observations and Data Reduction} \label{sect:data}
\galname, at a redshift of $z\approx 6.87$, is observed in the lensing field of Abell 2744, showing a Ly$\alpha$ absorption feature along with weak Ly$\alpha$ emission. Abell 2744 is a galaxy cluster at $z=0.308$ that has been extensively studied through multiwavelength observations. The abundance of archival data, particularly the recent JWST imaging and spectroscopic observations, combined with the magnification effect of the galaxy cluster, enhances our ability to investigate the properties of \galname. 

Inspired by the previously discussed GS9422, we aim to identify analogous galaxies in Abell 2744 that exhibit both Ly$\alpha$ emission and a DLA feature. We selected \galname as a candidate using the DAWN JWST Archive (DJA), a repository of public JWST galaxy data reduced with \grizli \citep{grizli} and \msaexp \citep{msaexp}.

\subsection{JWST/NIRCam and HST photometry} \label{subsect:nircam}
The mosaics for the Abell 2744 NIRCam imaging were processed by the UNCOVER project DR2\footnote{\url{https://jwst-uncover.github.io/DR2.html\#Mosaics}} 
\citep{image_data}.
A detailed description of the image reduction process is provided in \citet{image_data}. We included data from 7 filters in this research: F115W, F150W, F200W, F277W, F356W, F410M, and F444W. The data were produced using the JWST pipeline version 1.8.4, with the calibration file {\sc jwst 0995.pmap}. We displayed the images from JWST/NIRCam in the first column of Figure \ref{fig:galfits}. 

\begin{figure*}
 \centering
 \includegraphics[width=0.9\textwidth]{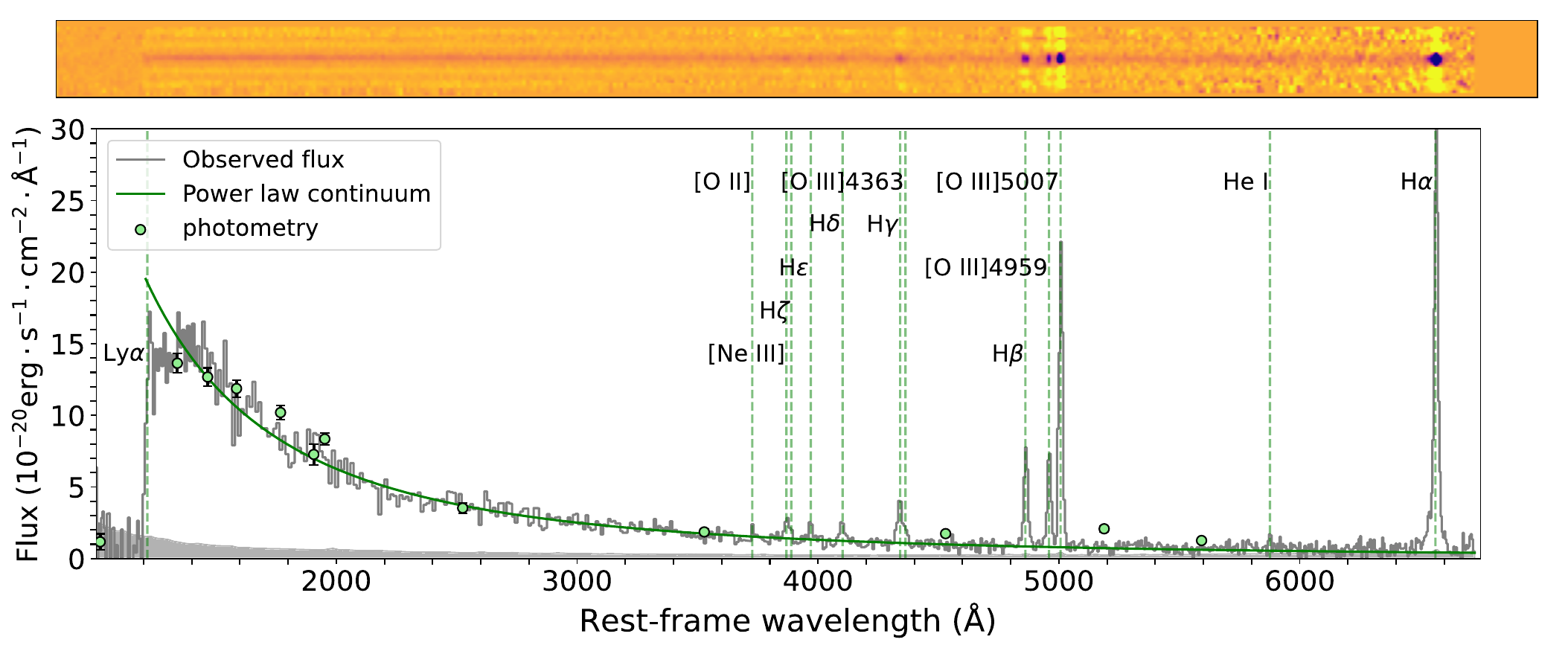}
 \vspace*{-1em}
 \caption{Images and rest frame spectroscopy of \galname.
\textbf{Top:} The 2D prism spectrum extracted using the \msaexp pipeline from Gabe Brammer.
\textbf{Bottom:} The extracted 1D spectrum is shown in gray, with the $1\sigma$ error indicated by the light gray shading. Broadband photometry measurements from the literature \citep{paris_catalog} are included as light green dots with error bars. The light green curve represents the best-fit power law model of the rest-frame UV continuum, spanning $1500-2600\ \mathrm{\AA}$. Important emission line wavelengths are marked with vertical dashed lines. It is evident that \galname exhibits a series of prominent Balmer lines and weak Ly$\alpha$ emission.}
\label{fig:spec}
\end{figure*}

To investigate this further, we compute the color indices
$\mathrm{F277W{-}F444W}$, $\mathrm{F277W{-}F356W}$, and
$\mathrm{F150W{-}F200W}$ in AB magnitudes using the photometric catalog
from \cite{paris_catalog}, and compare them with the LRD candidates \citep{LRDcandidate} in Figure~\ref{fig:color selection}. The color of \galname\ is not as red as that of the LRDs and it does not display a V-shaped SED. A more detailed discussion is provided in Section~\ref{sec:color_discuss}. The photometric data from \citet{paris_catalog}, including the ACS/WFC
(F814W) and WFC3/IR (F105W, F125W, F140W, and F160W) bands, together with the seven JWST/NIRCam bands mentioned above, are shown in Figure~\ref{fig:spec}.

\subsection{JWST/PRISM spectrum} \label{subsect:nirspec}
We utilize the low-resolution JWST/NIRSpec PRISM spectra from the UNCOVER project (DR4)\footnote{\url{https://jwst-uncover.github.io/DR4.html\#Spectra}} \citep[PIs: Labb\'e \& Bezanson; PID: GO-2561;][]{bezansonJWSTUNCOVERTreasury2024}, to study the continuum and emission line characteristics of \galname \citep{uncover_prism}. 
The observations combine NIRCam imaging (reaching depths of 29.5 to 30 AB magnitude in 8 filters) and low-resolution ($R \approx 100$) NIRSpec/PRISM spectroscopy (19-hour exposures), using strong lensing to probe faint galaxies at redshifts $z>10$ and during the Epoch of Reionization.
We use \msaexp to do reductions, which still use parts of the STScI pipeline but there are some changes included to increase the data quality \citep{msaexp}.

\section{Analysis and Results} \label{sect:rslt}
\subsection{Morphology fitting} \label{galfits}
\begin{figure}
 \includegraphics[width=0.5\textwidth]{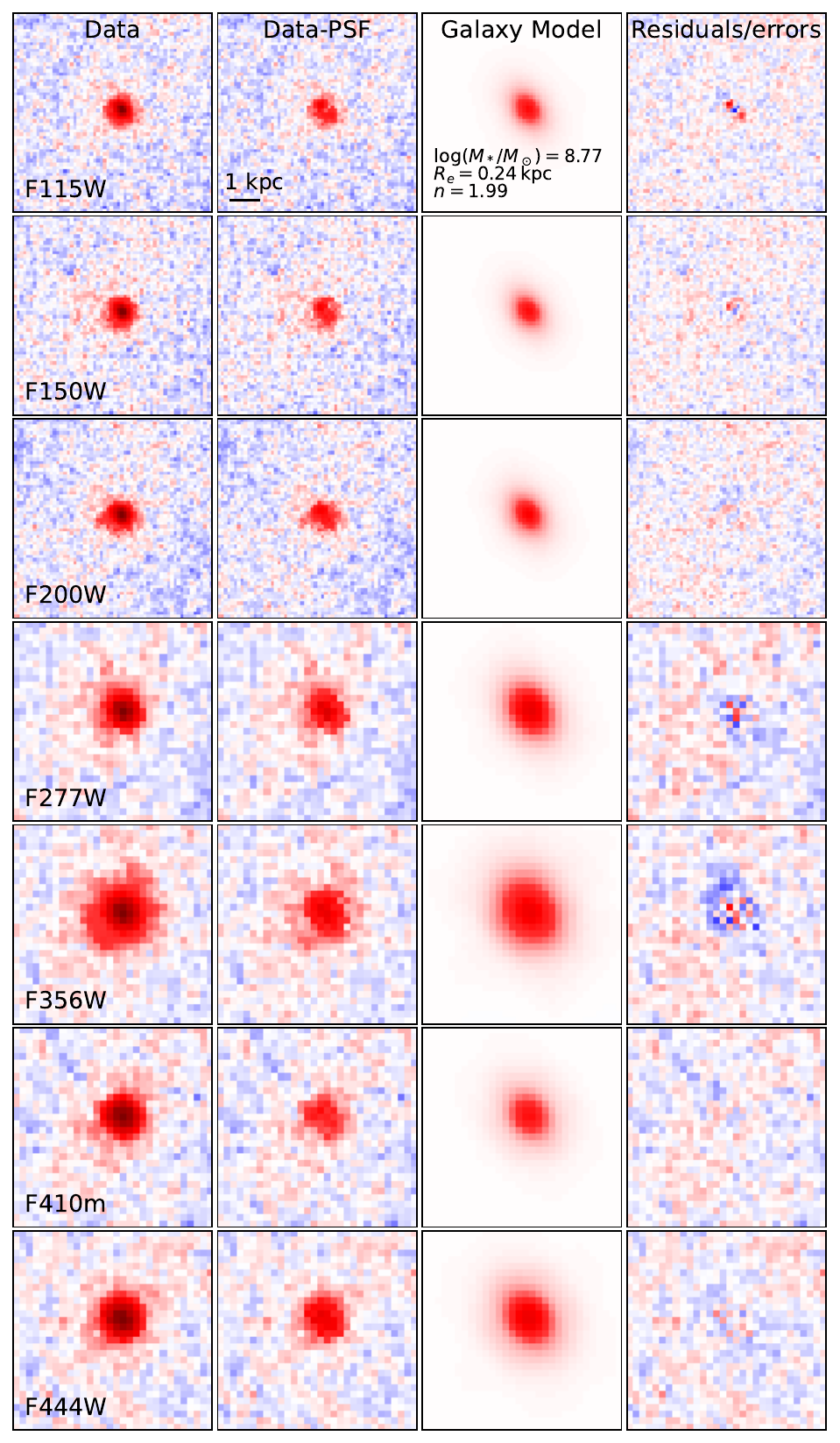}
 \vspace*{-1em}
 \caption{\galfits AGN–host decomposition of \galname. The first column displays the observed images in seven filters; 
 the second column shows data subtracted by PSF model; the third column presents the S\'{e}rsic model; and the fourth column shows the residuals divided by the errors. 
 We wrote the stellar mass, effective radius and S\'{e}rsic index on the lower left of the panel in first row. The residual images indicate that the model successfully reproduces the observed characteristics of \galname in most bands.}
\label{fig:galfits}
\end{figure}


We use \galfits\ \citep{galfits} to analyze the imaging data. \galfits\ is a recently developed multiband forward-modeling code that 
performs self-consistent image decomposition together with SED modeling across multiple bands. 
This approach allows us to separate the AGN and host contributions even in very compact sources, such as LRDs \citep{decompose1, decompose2, decompose4}.

An empirical Point Spread Function (PSF) model is constructed using the PSFEx \citep{PSFEx} program. The procedure begins by running SExtractor \citep{sextractor} on our science mosaic images to generate a source catalog, which serves as the input for PSFEx. PSFEx builds the PSF model by selecting a sample of bright, unsaturated, and isolated point sources from the catalog. 

This empirical approach is essential, as opposed to using a theoretical model. Our final science products are drizzled mosaic images with a refined pixel scale of $0.02''$/pixel (for short-wavelength band). In contrast, theoretical PSFs from tools like WebbPSF are typically simulated based on the native detector pixel scale ($0.031''$/pixel) This significant mismatch in sampling makes it inappropriate to directly apply a theoretical model to our SED fitting.

We tested two models in \galfits: a double Sérsic model and a Sérsic+PSF model. The Sérsic+PSF model provided a better fit to the observed imaging data. In this model, the AGN component is represented by a PSF \citep{galfits}, while the galaxy component is modeled with a single Sérsic profile. In addition to the Sérsic profile parameters, we treated the dust extinction (\(A_V\)), stellar mass, and the nebular ionization parameter (\(\log \rm U\)) as free parameters during the fitting process. No priors were applied to the AGN continuum, meaning the emission from the PSF component remains arbitrary.

The fitting results are shown in Figure~\ref{fig:galfits}. It is evident that the S\'{e}rsic+PSF model successfully reproduces the observed features of \galname\ across most bands. 

According to the results from \galfits, the galaxy component dominates all seven bands. 
The PSF component contributes significantly to the central region of \galname, especially in the long wavelength bands (F277W, F356W, F410M, and F444W) as shown in the second and third column of Figure~\ref{fig:galfits}. We emphasize that the central PSF cannot be entirely attributed to the AGN in the \galfits modeling. A compact central stellar component could also contribute to the PSF in this region.

We noticed that the model does not fit very well in the F115W, F277W, and F356W bands. In F115W, there may be some overestimation of the PSF model, leading to negative residuals at the center. The images in F277W and F356W exhibit complex structure, suggesting that a single S\'{e}rsic profile may not fully describe the host galaxy, which requires further investigation.

\subsection{SED fitting} \label{subsect:sed}
With the help of \bagpipes \citep{bagpipes1, bagpipes2}, we perform broadband spectral energy distribution (SED) fitting of our photometric data from \citet{paris_catalog} and spectroscopic data. Our basic assumptions include an exponentially declining star formation history (SFH) and the Calzetti dust extinction law \citep{Calzetti2000}, with the visual extinction $A_{\rm V}$ ranging from 0 to 1.5. In this model, the e-folding timescale parameter $\tau$ spans from 0.3 to 10 Gyr, and the stellar metallicity ($Z/Z_\odot$) ranges from 0 to 2.5. The integrated fitting results for stellar mass $\log(M_{\ast}/M_{\odot}) = 8.77 ^{+0.24}_{-0.11}$. The star formation rate, metallicity, and stellar age are presented in Table \ref{tab:gal}. We also measure the UV slope, $\beta = -2.46^{+0.03}_{-0.03}$, 
by fitting the SED model spectrum over the rest-frame wavelength range of $1500\,\mathrm{\AA}$ to $2600\,\mathrm{\AA}$. The result is consistent with that obtained from a simple power-law fit to the observed spectrum.
 
We note that \bagpipes does not include AGN components, while \galname exhibits a broadened H$\alpha$ emission line, a feature commonly associated with AGN activity. To account for this, we further perform SED fitting using \cigale \citep{cigale}, which incorporates the SKIRTOR2016 module \citep{skirtor2016_2,skirtor2016}. However, because this AGN module does not include AGN emission lines, the broad component of the H$\alpha$ emission is not modeled. The AGN fractions inferred from \cigale are substantially smaller than the PSF-component fractions derived from \galfits. In F277W, F356W, and F444W, the AGN fractions are 0.07, 0.06, and 0.10, respectively. Therefore, we infer that the PSF component in the GalfitS decomposition is predominantly contributed by stellar emission. 

We also note that \cigale predicts a very strong Ly$\alpha$ emission line. Given the UV and optical continua, the nebular model appears to favor such a strong Ly$\alpha$ emission, which is not observed in the data.

\begin{figure*}
     \centering
    \includegraphics[width=0.9\textwidth]{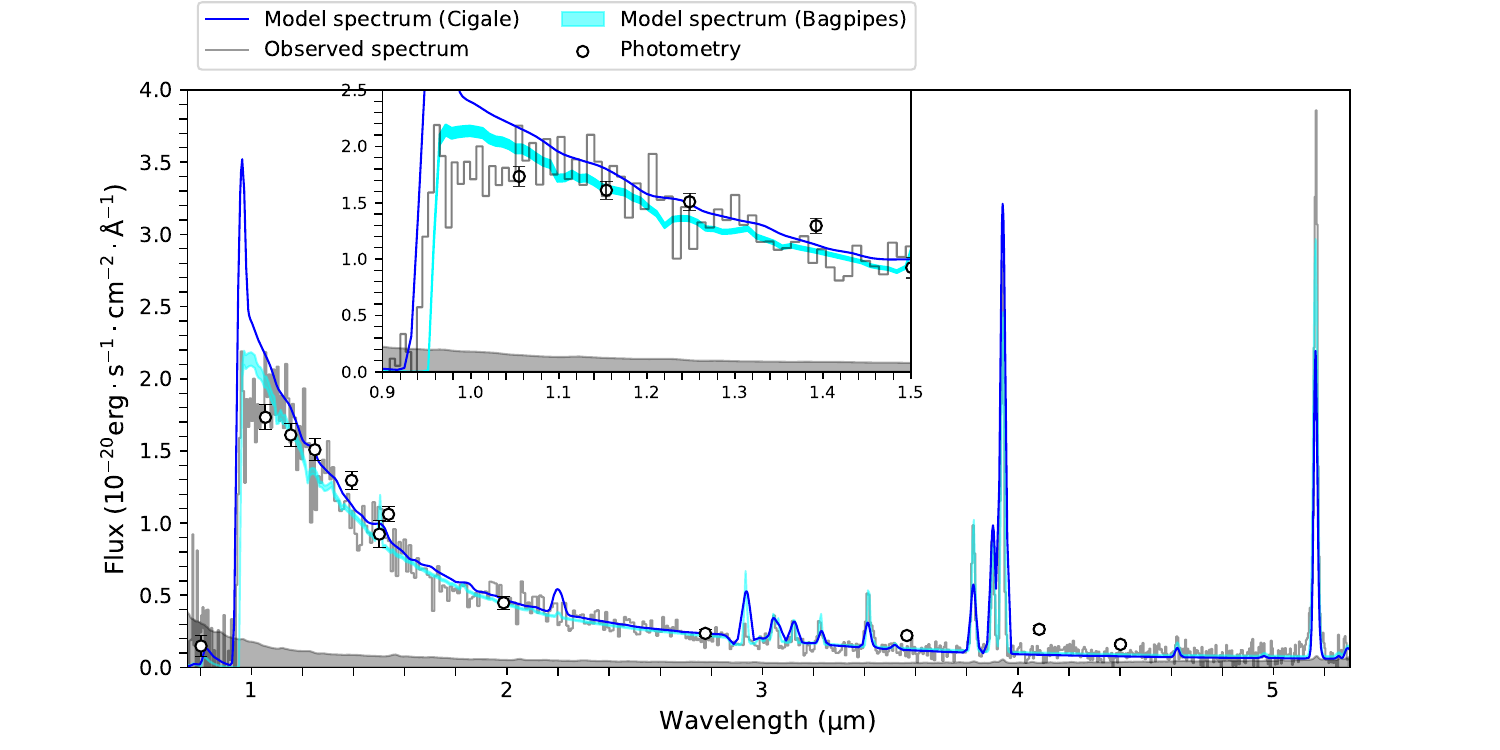}
    \vspace*{-1em}
   \caption{SED fitting results from \cigale and \bagpipes.
The total model spectra from \cigale\ and \bagpipes are shown as blue and cyan curve. The observed photometric fluxes for the filters F814W, F105W, F125W, F140W, F160W, F115W, F150W, F200W, F277W, F356W, F410M, and F444W are shown as white points. The observed flux and error are represented by the gray curve and shaded region, respectively. We zoom in on the UV turnover part of the spectrum in the upper-right corner of this figure. 
Both the \cigale\ and \bagpipes\ models reproduce the UV and optical observations well, but neither fits the curvature at the red end of 1216$\ \mathrm{\AA}$ successfully.
}
\label{fig:AGN_SED}
\end{figure*}

From \cigale, the best-fit stellar mass is $\log(M_{\ast}/M_{\odot}) = 8.59 \pm 0.15$, consistent with the result obtained from \bagpipes.  
The SED models generated by these codes are presented in Figure~\ref{fig:AGN_SED}. The figure shows that, at least in the UV and optical bands, the continuum is dominated by star formation rather than AGN emission, which is consistent with the results from \galfits.



\subsection{Emission line fitting} \label{subsect:EL}
To study the physical properties of the rest-frame optical emission lines, we first fit the continuum 
near each line with fifth-order polynomial and subtract it from the observed spectrum to isolate the emission features. We then use \lmfit \citep{lmfit}, to model several individual emission lines with Gaussian profiles. In this analysis, the intrinsic velocity dispersions of 
H$\beta$, H$\gamma$, H$\epsilon$, and H$\delta$ are constrained to match the value of the narrow H$\alpha$ component. Similarly, the intrinsic velocity dispersions of [O \III] $\lambda4363$ and [O \III] $\lambda4959$ are set to that of [O \III] $\lambda5007$, while the flux ratio between [O \III] $\lambda5007$ and [O \III] $\lambda4959$ is fixed at 2.98. Additionally, the velocity offset of [O \III] $\lambda4363$ and [O \III] $\lambda4959$ is anchored to [O \III] $\lambda5007$. We account for the line-spread function (LSF\footnote{We use the fiducial resolving power curves from \url{https://jwst-docs.stsci.edu/jwst-near-infrared-spectrograph/nirspec-instrumentation/nirspec-dispersers-and-filters}}) by convolving our model spectrum with Gaussians whose resolution varies with wavelength.

\begin{figure*}
 \centering
 \includegraphics[width=0.9\textwidth]{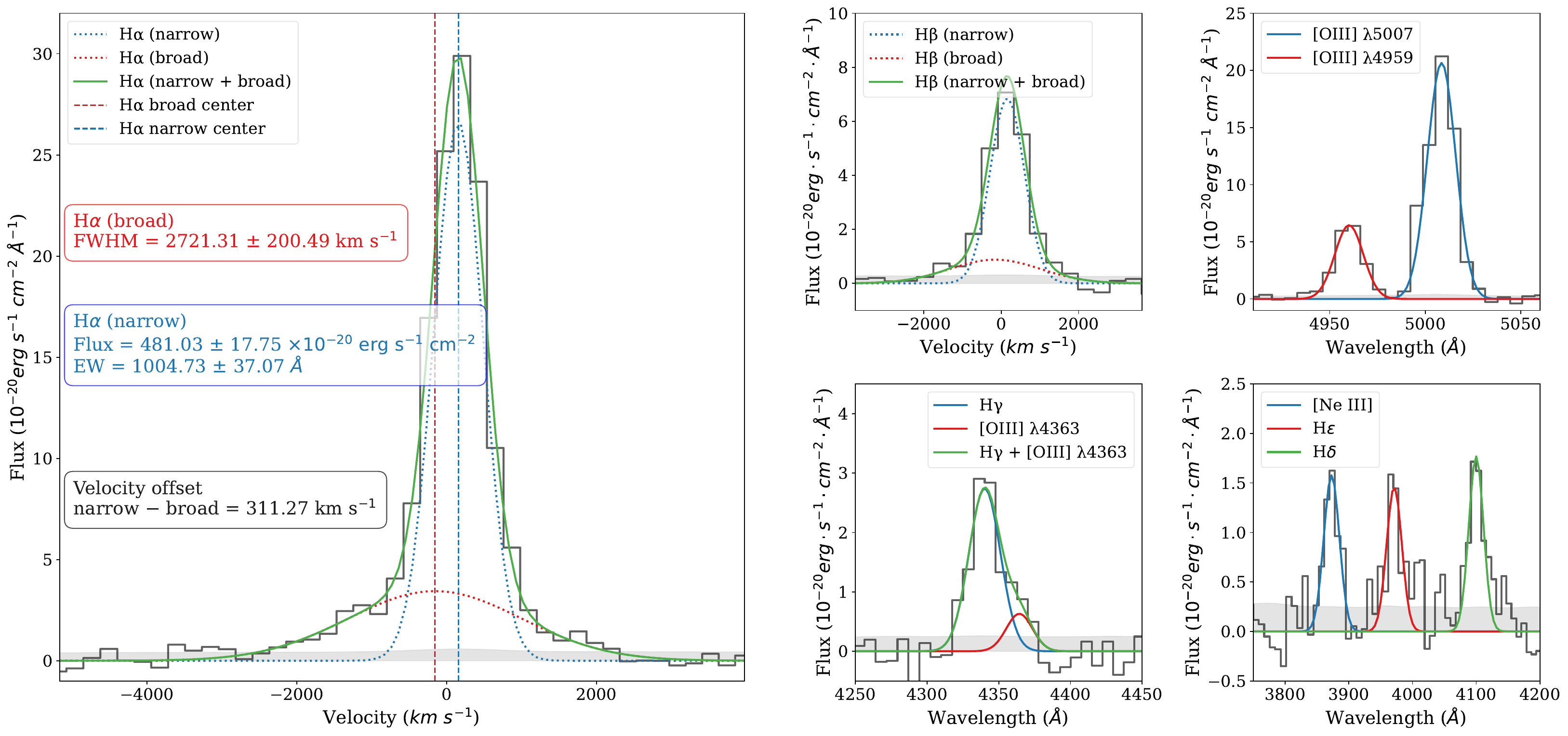}
 \vspace*{-1em}
 \caption{\small
Fitting results of optical emission lines identified in the JWST/NIRSpec spectrum of \galname. The panels show $F_\lambda$ (in units of $10^{-20} \mathrm{erg}\cdot\mathrm{s}^{-1}\cdot\mathrm{cm}^{-2}\cdot\mathrm{\AA}^{-1}$) as a function of velocity offset or rest-frame wavelength. In the left panel, we showed more detailed information about the two-component Gaussian fitting of H$\alpha$ emission. Relevant measurements, such as line flux and FWHM, are displayed in this panel, using the same color as the corresponding best-fit model.
All results presented here are before correction for dust extinction and lensing magnification. Emission-line measurements are summarized in Table \ref{tab:ELflux}. A detailed description of the fitting method and a discussion of line ratios are provided in Sections \ref{subsect:EL} and \ref{subsect:ratio}.
\label{fig:lines}}
\end{figure*}

From Figure~\ref{fig:lines}, it is evident that the H$\alpha$ emission line is broadened. We model this line with a two-component Gaussian profile and find that including a broad component significantly improves the fit. To test the robustness of this result, we computed the Bayesian Information Criterion (BIC) for both models: $BIC_{\rm single} = 176.51$ and $BIC_{\rm two-component} = 63.74$. The lower BIC value for the two-component model confirms that the broad H$\alpha$ component is real rather than an artifact of overfitting. However, there is not detected broad component in [O \III] $\lambda$5007 and [O \III] $\lambda4959$. We attempted to include a second Gaussian component in the fitting process of [O III] $\lambda5007$ line. The additional component does not improve the fit, and the BIC value becomes even larger, which indicates that there is no evidence for an outflow-related wing in [O III].

A broad H$\alpha$ emission component is commonly observed in Type I AGN. Although the spectrum of \galname\ lacks sufficient high-ionization lines to definitively confirm its AGN nature, this remains a plausible scenario. If the observed broad H$\alpha$ emission arises from gas kinematics in the broad-line region \citep[e.g.,][]{BM1,BM2,BM3}, and given the well-established correlations between the luminosities of Balmer lines and the continuum luminosity at $5100\,\mathrm{\AA}$ ($L_{5100} = \lambda L_\lambda$ at $\lambda = 5100\ \mathrm{\AA}$), the black hole mass can be estimated using the broad H$\alpha$ component following Equation (6) of \citep{BM1}.
\begin{equation}
\begin{split}
    M_{\mathrm{BH}}  =(2_{-0.3}^{+0.4})\times10^{6}\left(\frac{L_{\mathrm{H\alpha}}}{10^{42}\mathrm{ergs~s^{-1}}}\right)^{0.55\pm0.02} \\
\times \left(\frac{\mathrm{FWHM_{H\alpha}}}{10^{3}\mathrm{km~s^{-1}}}\right)^{2.06\pm0.06}M_{\odot}. 
\end{split}
\end{equation}\label{BH_eq}
However, we caution that the broadening of permitted lines may also arise from electron scattering in dense ionized gas, producing exponential rather than Gaussian line wings \citep{resonant_scattering}, which may lead to an overestimate of the black hole mass. Distinguishing between these mechanisms requires deep, high-resolution spectroscopy, which is beyond the scope of this work.

The luminosity and FWHM of the broad component H$\alpha$ were substituted into the equation above, yielding 
$M_{\mathrm{BH}}=2.90^{+2.35}_{-1.28}\times 10^7 M_\odot$ and a black hole–to–stellar mass ratio of $\log (M_{\rm BH}/M_{\rm stellar}) = -1.58^{+0.45}_{-0.34}$, which is typical for a supermassive black hole. It should be carefully considered that gas kinematics may not satisfy all the conditions of the standard AGN model, and some studies suggest that the masses might be overestimated \citep{over_mass1,black_hole_star,over_mass3}. However, direct dynamical measurement of black hole mass in a lensed LRD at $z = 7.04$ yields a conservative estimate of $M_{\rm BH}/M_{\rm stellar} > 2$, suggesting the presence of an almost ``naked" black hole \citep{direct_measurement}.

We attempted to place an upper limit on the broad component of H$\beta$ by tying both the narrow and broad components to those of H$\alpha$. Although the broad H$\beta$ component is not as prominent as in H$\alpha$, the fit is indeed improved. We then applied the formalism from \cite{BHmass} using the broad H$\beta$ component and the monochromatic luminosity at $5100\,\mathrm{\AA}$. The resulting black hole masses are $\log M_{\rm BH} \approx 8.98$, $8.51$, and $8.84$ for classical bulges, pseudobulges, and the combined sample, respectively—values that are extremely large compared to the stellar mass. Given that the system is more likely stellar-dominated, we do not consider this a meaningful constraint.

We derived a dust extinction of $\rm A_V = 1.15 \pm 0.23$ using the narrow components of both H$\alpha$ and H$\beta$. Estimating $\rm A_V$ from the Balmer decrement in AGN systems can be problematic: \citet{AGN_dust_20, AGN_dust_velocity} found an anti-correlation between H$\alpha$/H$\beta$ and continuum or line flux in a 20-year optical monitoring campaign of the Seyfert~1 galaxy NGC~7603. CLOUDY simulations, theoretical recombination spectra, and SDSS observations consistently suggest that the Balmer decrement is reliable only for BLR regions with high ionization parameter or density at large column densities \citep{AGN_dust_phy}. To minimize these effects, we use only the narrow Balmer components, ensuring that our estimate is not driven by BLR emission. However, since we lack evidence for long-term variability in \galname, this result should be interpreted with caution.

We also attempted to place an upper limit on the [N \II] $\lambda\lambda 6548, 6584$ lines by subtracting the best-fit H$\alpha$ Gaussian model from the continuum-subtracted spectrum. We then fixed the line width to that of the narrow H$\alpha$ component and fitted the residuals. Due to the limited spectral resolution, we adopted a single Gaussian model for the fit. For the [O II] $\lambda\lambda3726,3729$ doublet, we estimate the upper limits of their fluxes by fixing the line width to that of [O III] $\lambda5007$, as no reliable detection of these lines is present in the observed spectrum. Similarly, we derive the upper limit for He I by fixing the intrinsic line width to that of the narrow H$\alpha$ component and performing a Gaussian fit. Representative fitting results are shown in Figure \ref{fig:lines} and summarized in Table \ref{tab:ELflux}.

\subsection{UV continuum and $Ly\alpha$ emission} \label{subsect:cont}
\galname exhibits a strong sign of UV turnover, which is commonly observed in DLA systems. However, inspired by the recently widely discussed galaxy GS9422 at $z=5.943$ \citep{nebular_cont,GS9422_1,GS9422_2,GS9422_3}, we considered the nebular continuum as a possible source of the UV continuum in \galname. 

For the potential DLA system, we use the same approach as \cite{DLA_fit} to construct the model.
We modeled the intrinsic continuum with a simple power law fitted over the rest-frame wavelength range of $1500\,\mathrm{\AA}$ to $2600\,\mathrm{\AA}$. We assume that the effect of residual neutral hydrogen within the ionized bubble can be neglected.

The modeling of the DLA system includes the effects of HI column density, velocity offset and covering fraction. The sightline that is covered by the DLA are also taken into account \citep{DLA_fit}. The absorbed spectra resemble the Voigt-Hjerting function, and we use the algorithms described in \citet{voigt_function} to approximate it.

The remaining photons interact with the neutral Intergalactic Medium (IGM). The foreground neutral hydrogen absorbs photons that have been redshifted to $1216\ \mathrm{\AA}$ relative to them. 
We use the same method as \citet{IGM_function} to model the IGM absorption, incorporating necessary corrections based on the Gunn-Peterson effect to calculate the optical depth of Ly$\alpha$ \citep{strong_damped}.

Since the observed spectrum of \galname shows an inconspicuous Ly$\alpha$ emission line, we add this component to our model. We use a Skew-Normal Function for the Ly$\alpha$ profile to remove the emission on the blue side. 

During the fitting process, we convolve the model spectrum with Gaussians of variable resolution to account for the effect of the LSF. The resolution curve adopted here is the same as that used in Section~\ref{subsect:EL}. We treated the HI column density of the DLA ($\rm N_{HI}$, reported in logarithmic form), the velocity offset of the DLA relative to the target galaxy ($\Delta v$), the equivalent width of the Ly$\alpha$ emission line ($\rm EW_{Ly\alpha}$, reported in logarithmic form), the covering fraction ($f_c$), and the size of the bubble ($\rm R_{ion}$) as free parameters. In this model, $\log (\rm N_{HI}/{\rm cm^{-2}})$ ranges from 21 to 24, $\Delta v$ ranges from $-50$ to $0\ \times 10^4\ \mathrm{km\ s^{-1}}$, $\log (\rm EW_{Ly\alpha}/{\rm \AA})$ ranges from $-3$ to $3$, $f_c$ ranges from 0 to 1, and $\rm R_{ion}$ ranges from 0.5 to 2 Mpc. The best-fitting values are $\log (\rm EW_{Ly\alpha}/{\mathrm{\AA}}) = 1.89_{-0.07}^{+0.08}$, $\log (\rm N_{HI}/{\rm cm^{-2}}) = 22.20_{-0.37}^{+0.32}$, $\Delta v = -1.50_{-0.31}^{+0.32}\times 10^4\ \mathrm{km\ s^{-1}}$, $f_c = 0.24_{-0.05}^{+0.08}$, and $\rm R_{ion}=1.17^{+0.56}_{-0.47}$ Mpc. The corresponding Ly$\alpha$ flux is measured to be $84.61^{+110.97}_{-41.21} \times 10^{-20}\ \mathrm{erg\ s^{-1}\ cm^{-2}}$. All uncertainties presented here correspond to the $1\sigma$ confidence intervals. A full summary of the fitting results obtained from these models is provided in Table~\ref{tab:gal}. The observed absorption profile suggests a very large inflow velocity based on the best-fit velocity offset; however, we caution that this may be affected by the low spectral resolution.



Since the Ly$\alpha$ emission is so weak, we must rule out the possibility that it is a defective pixel. Therefore, we masked the wavelength range from $1210\ \mathrm{\AA}$ to $1230\ \mathrm{\AA}$ and repeated the above process. It turns out that the best-fit equivalent width of the Ly$\alpha$ emission line becomes smaller, meanwhile the HI column density of the DLA and the cover fraction become larger. Thus, the observed Ly$\alpha$ emission is real if we adopt this model.

\begin{figure*}
 \centering
 \includegraphics[width=0.8\textwidth]{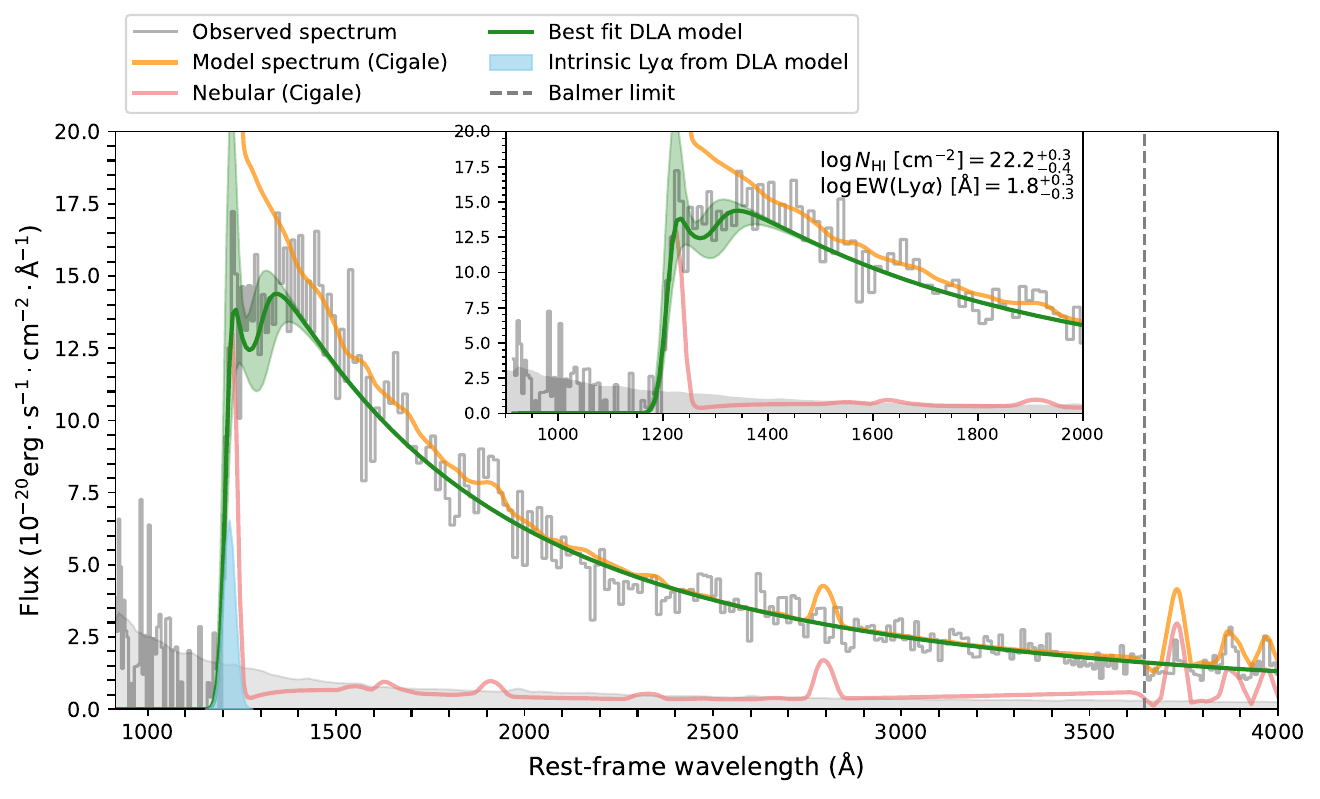}
 \vspace*{-1em}
 \caption{
Observed spectroscopic data and best-fit models. The rest-frame NIRSpec/prism 1D spectra and their $1\sigma$ uncertainties are shown as gray lines and shaded regions, respectively. The best-fit DLA models and their uncertainties are plotted as green lines and shaded regions. The lower limit of intrinsic Ly$\alpha$ estimated from the DLA model is shaded in sky blue. The model spectrum and nebular components from the \cigale\ are plotted in orange and pink, respectively. We also mark the Balmer limit with black vertical lines. We zoomed in on the UV turnover part of the spectrum and wrote the best-fitting H\I \ column density and Ly$\alpha$ flux in the upper-right corner of this figure. Even considering the nebular emission, it is difficult to reproduce both the UV curvature and the steep UV slope $\beta$ simultaneously, so the DLA absorption is required.
}
\label{fig:continuum}
\end{figure*}

In order to rule out the possibility that nebular emission, especially the two-photon emission, dominates the UV continuum turnover instead of DLA, we plotted the model spectrum from the \cigale\ SED fitting in Figure \ref{fig:continuum} for comparison. However, without DLA, this model fails to reproduce the UV continuum well. The model spectrum appears much higher than the observed data at the red end of the 1216\AA\ and exhibits a very strong Ly$\alpha$ emission line.

The fitting results of both the DLA and stellar+nebular emission models are plotted in Figure \ref{fig:continuum} with green and orange lines, respectively. The green shaded area corresponds to the $1\sigma$ uncertainty of the best-fit DLA model. Furthermore, we plot the pure Ly$\alpha$ emission line without the continuum, derived from the DLA model.

{\small
\begin{table*}
\centering
\begin{threeparttable}
\caption{Rest-frame optical emission line properties of \galname.\label{tab:ELflux}}
\begin{tabular}{lcccc}
 \hline\hline 
 Emission Line  &  Observed Flux$^a$  & extinction-corrected flux$^b$     &  Rest-frame Equivalent Width \\
 &[$10^{-20}erg\cdot s^{-1}cm^{-2}$]  &[$10^{-20}erg\cdot s^{-1}cm^{-2}$] & [$\mathrm{\AA}$]\\
 \hline \noalign {\smallskip}
 
H$\alpha$ narrow & $481.03\pm17.75$ & $1175.96^{+279.22}_{-237.80}$ & $1004.73\pm37.07$ \\ [\narrow]

H$\alpha$ broad & $218.19\pm31.54$ & $533.80^{+195.44}_{-155.61}$ & $443.57\pm64.13$ \\ [\narrow]

[O \III] $\lambda 5007$ & $383.73\pm21.56$ & $1232.83^{+407.38}_{-322.90}$ & $425.22\pm23.89$ \\ [\narrow]

[O \III] $\lambda 4959$ & $128.77\pm7.24$ & $418.47^{+139.55}_{-110.35}$ & $141.71\pm7.96$ \\ [\narrow]

H$\beta$ narrow & $123.48\pm6.31$ & $411.25^{+137.15}_{-108.39}$ & $180.73\pm0.51$ \\ [\narrow]

H$\beta$ broad & $41.91\pm12.91$ & $139.59^{+92.04}_{-64.62}$ & $62.06\pm0.33$ \\ [\narrow]

[O \III] $\lambda 4363$ & $14.45\pm5.16$ & $55.20^{+42.45}_{-28.44}$ & $12.82\pm 4.58$ \\ [\narrow]

H$\gamma$ & $75.33 \pm 5.64$ & $289.60^{+116.67}_{-87.85}$ & $66.49\pm 4.98$ \\ [\narrow]

H$\delta$ & $51.36\pm 6.04$ & $211.12^{+100.93}_{-72.80}$ & $62.18\pm7.31$ \\ [\narrow]

H$\epsilon$ & $43.72\pm6.54$ & $185.80^{+98.56}_{-69.31}$ & $46.89\pm7.01$ \\ [\narrow]

[Ne \III] $\lambda 3869$ & $48.42 \pm 6.53$ & $210.52^{+108.98}_{-76.88}$ & $39.95\pm5.39$ \\ [\narrow]

[O \II] $\lambda\lambda 3726,3729$ & $<30.66$ & $<136.98$ & - \\ [\narrow]

[N \II] $\lambda\lambda 6548,6584$ & $<10.02$ & $<29.13$ & - \\ [\narrow]

He \I\ $\lambda 5877$ & $<20.50$ & $<55.682$ & - \\
[\narrow]

 \hline \noalign {\smallskip}
 \end{tabular}
    \begin{tablenotes}
      \item[NOTE ---] The observed line fluxes and equivalent widths are measured from our emission line analyses of the NIRSpec/MSA prism spectrum with the \lmfit software (see \ref{fig:lines} for details). 
      \item[$a$] The observed line fluxes and upper limits before the corrections of dust extinction and lensing magnification.
      \item[$b$] The intrinsic line fluxes and upper limits after the corrections of dust extinction. We adopt $A_{\rm V}=1.15\pm0.23$ measured from the Balmer decrement ($\rm H\alpha/H\beta$).
    \end{tablenotes}
\end{threeparttable}
\end{table*}

}


\section{Discussion}\label{sec:discuss}

\subsection{HI column density and DLA system}
DLA systems typically contain high column density HI absorbers, which block the escape of Ly$\alpha$ photons \citep{reviewDLA,introDLA}. However, several DLA systems, such as \galname, exhibit weak Ly$\alpha$ emissions. Previous studies during the Cosmic Noon have shown that Lyman-alpha emitters (LAEs) tend to cluster around absorbers with high HI column densities. This seemingly contradictory scenario arises from the complex structure of the circumgalactic medium (CGM), which is patchy and inhomogeneous \citep{lya_intro1,HIdensity}.

Additionally, Ly$\alpha$ emitters during cosmic noon are relatively small compared to galaxies with similar UV luminosities \citep{compact_lya}. This suggests that a compact morphology plays a crucial role in facilitating the escape of Ly$\alpha$ photons. Since the size-luminosity relation at redshift $z\sim 3-4$ is consistent with that of local Green Pea galaxies, the trend of LAEs being more compact does not evolve with redshift. \galname, a Ly$\alpha$ emitter during the reionization epoch, features a compact size and outflows. This object can help us extend our understanding of this relationship to higher redshifts and potentially uncover the underlying mechanisms.

\galname shows both DLA and Ly$\alpha$ emission. We fit the continuum with two models. One incorporates both DLA, Ly$\alpha$ emission and IGM absorption, the other is considered a nebular continuum, Ly$\alpha$ emission and IGM absorption. It appears that the former can reproduce the UV spectrum better. 
We find that the DLA model does not constrain the ionized bubble size well, indicating that the UV turnover is dominated by nearby neutral H I outside the bubble and is therefore insensitive to $\rm R_{ion}$.
However, \galname exhibits a very faint Balmer jump, resembling a feature of free-bound emission, which typically dominates the nebular continuum. Therefore, we cannot rule out the contribution of nebular emission to the continuum. 

However, it is important to note that high resolution data is required to better distinguish the effect of DLA versus IGM when modeling the absorber and estimating the intergalactic neutral hydrogen fraction. 
As discussed by \cite{pitfall}, degeneracies between the DLA column density ($\rm N_{DLA}$) and the neutral fraction of the IGM ($x_{\rm HI}$) can compromise the accuracy of such measurements. In addition, when only a single absorption component is fitted, low-resolution spectra may result in an overestimation of $N_{\rm DLA}$ \citep{overestimate}.

\subsection{AGN diagnostics}
\begin{figure}\label{fig:AGN_diag}
 \includegraphics[width=0.5\textwidth]{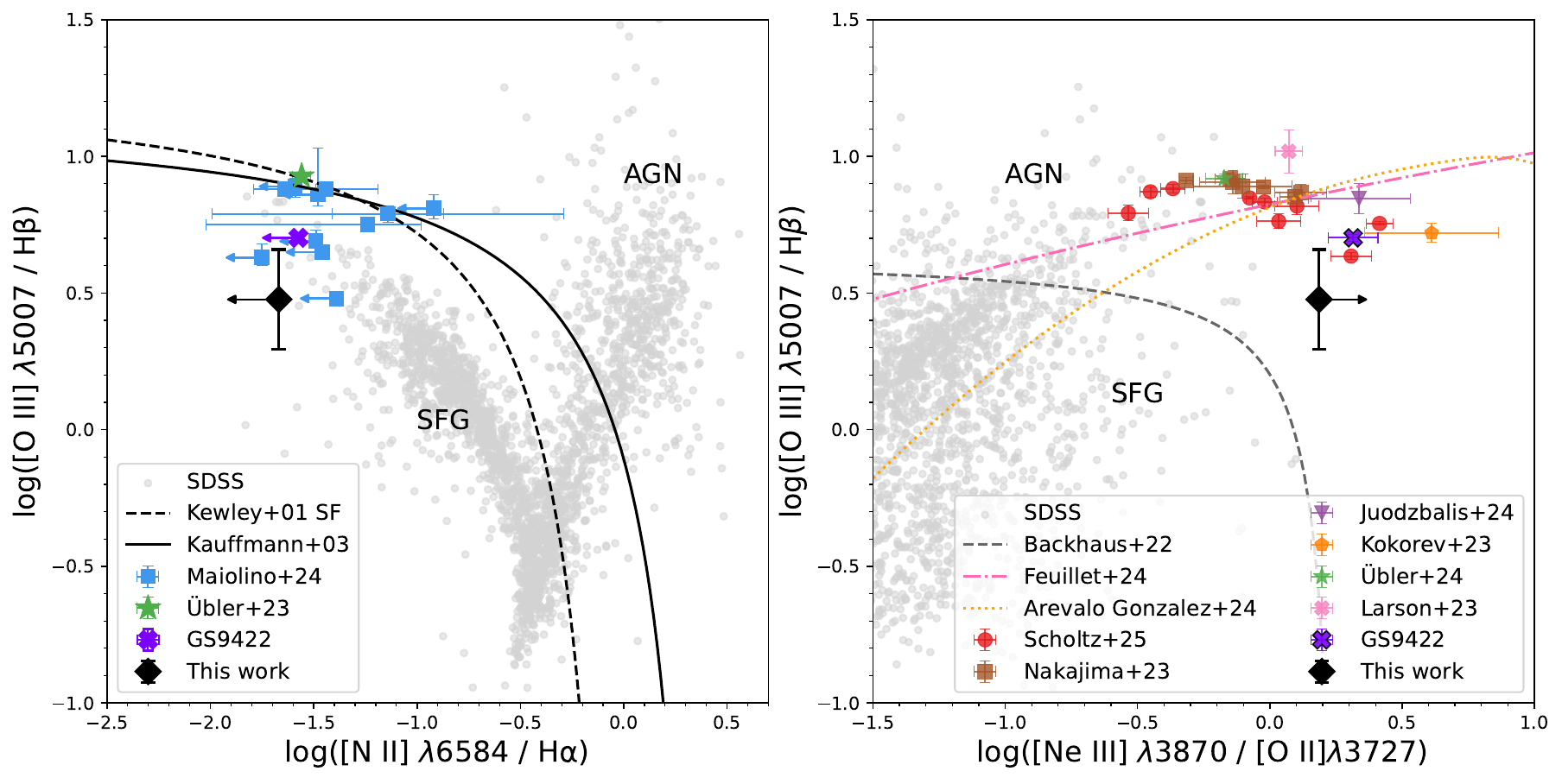}
 \vspace*{-1em}
 \caption{
 Emission-line diagnostics of star formation/AGN activity.
\textbf{Left:} 
N2–BPT diagram from \cite{BPT1} and \cite{BPT2}, along with comparison samples from \cite{maiolino+24} and \cite{ubler+23}, marked with purple squares and a green star. Emission lines of GS-9422 from \citet{nebular_cont} are shown with a purple cross. Our object is marked with a black diamond.
\textbf{Right:} The relation between [O \III] $\lambda 5007$/H$\beta$ and [Ne \III] $\lambda 3869$/[O \II] $\lambda\lambda 3727,3729$ (OHNO). We compare our object with narrow-line AGNs at high redshift \citep{scholtz+23}, galaxies at redshifts $z=4-10$ discovered by JWST \citep{nakajima+23}, high-redshift overmassive black holes and broad-line AGNs \citep{juodvzbalis+24, kokorev+23, ubler+24, BM3}, as well as GS9422 \citep{nebular_cont}. 
 }
\end{figure}

We plot AGN diagnostic diagrams to assess the likelihood that our object hosts an AGN. In the left panel of Figure~\ref{fig:AGN_diag}, we show the classical N2–BPT diagram from \cite{BPT1} and \cite{BPT2}, along with comparison samples from \cite{maiolino+24} and \cite{ubler+23}, as well as GS9422 from \citet{nebular_cont}. We adopt [O \III]~$\lambda5007$, H$\beta$, H$\alpha$, and [N \II]~$\lambda6583$ measurements from JWST/NIRSpec G395H/F290LP, while [Ne \III]~$\lambda3869$ and [O \II]~$\lambda\lambda3727,3729$ are taken from G235M/F170LP. 
\galname falls within the star-forming galaxy (SFG) region in this panel, although high-redshift AGN samples often lie near the boundary of this classification. The right panel presents the OHNO diagram, with demarcation curves from \cite{OHNO1}, \cite{OHNO2}, and \cite{OHNO3}. According to the \cite{OHNO1} criteria, our galaxy lies within the AGN region. All emission-line measurements used in this section correspond to the narrow components only.

However, some recent works suggest that commonly used AGN diagnostics may not distinguish star forming galaxies and accreting black holes very well. \cite{cannot_distinguish} suggest that AGN narrow-line regions and star-forming H II regions occupy strongly overlapping regions in all common strong-line diagnostic diagrams, and low-metallicity AGN naturally produce strong-line ratios that are indistinguishable from extreme star-forming populations. \cite{assess_diag} pointed that many traditional optical diagnostics such as [O \III]/H$\beta$ vs [N \II]/H$\alpha$ perform very poorly for dwarf galaxies. Therefore, we cannot rule out the possibility that the narrow lines of \galname originate from the AGN narrow-line region, or that they are dominated by the ISM in the host galaxy.

\subsection{Physical conditions}\label{subsect:ratio}
We derive the electron temperature ($T_e$) from the [O \III] $\lambda4363$ emission line.
We adopt the relation $R_{O3} = \frac{I(4959) + I(5007)}{I(4363)}$ and calculate $T_e([\mathrm{O}_{\III}]) = 0.7840 - 0.0001357R_{O3} + \frac{48.44}{R_{O3}}$, where $T_e$ is expressed in units of $10^4$ K \citep{Te_method}. 

This empirical relation assumes an electron density of $n_e = 100\ \mathrm{cm}^{-3}$ and is valid for $T_e$ in the range $0.7$ to $2.5 \times 10^4$ K. 
\galname\ exhibits a [O \III]\,$\lambda 4363$ emission line blended with H$\gamma$, resulting in a $T_e$ of $2.4 \times 10^4$ K. Using Equation (1) from \citet{metallicity}, we estimate the metallicity of \galname\ to be $12 + \log(\mathrm{O/H}) \approx 7.48$. Although these electron temperature and metallicity values may not be accurate for the low SNR of [O \III] $\lambda 4363$, 
this estimate still indicates the extremely high electron temperature and metal-poor narrow line region (NLR). 

Several studies of LRDs have reported exceptionally high $\mathrm{I}(4363)/\mathrm{I}(5007)$ ratios \citep{kokorev+23_diag, jones2025blackthunder}. According to \citet{jones2025blackthunder}, such elevated [O III] $\lambda4363$/[O III] $\lambda5007$ values may arise from a multiphase ISM—containing both diffuse and dense gas—or from extremely high electron temperatures ($T_e > 10^5\ \mathrm{K}$). Furthermore, the unusually large [O III] $\lambda4363$/H$\gamma$ ratio suggests that the AGN not only ionizes and overheats the gas, but that the host galaxy must also contain gas with very high densities. For the source discussed by \citet{jones2025blackthunder}, the required gas density is on the order of $10^{7}\ \mathrm{cm^{-3}}$.

\subsection{Comparison with red compact sources}\label{sec:color_discuss}
\subsubsection{Color Selection}
Inspired by the peculiarities of \galname, we compare it with other galaxies from the UNCOVER program \citep{uncover}, focusing on their color and morphology, particularly the LRDs candidates. We adopt the criteria outlined in \citet{LRDcandidate, LRDcandidate2} and \citet{LRD_discover2}. The goal of this selection is to identify objects that exhibit a red continuum slope in the rest-frame optical while also showing significant emission in the rest-UV, referred to as the ``V-shape" color selection.

We first apply a cut of $\rm{SNR_{F444W}} > 14$ and $m_{\rm{F444W}} < 27.7$ mag to select well-detected sources. An object is considered to have red color and compact size if it satisfies all of the following criteria:

$$red1 = \rm{(F115W - F150W < 0.8)} $$
$$
red2 = \rm{(F150W - F200W < 0.8)} $$
$$compact = \rm{f_{F444W}(0.4'')/f_{F444W}(0.2'') < 1.7}$$

\begin{figure*}
 \centering
 \includegraphics[width=0.8\textwidth]{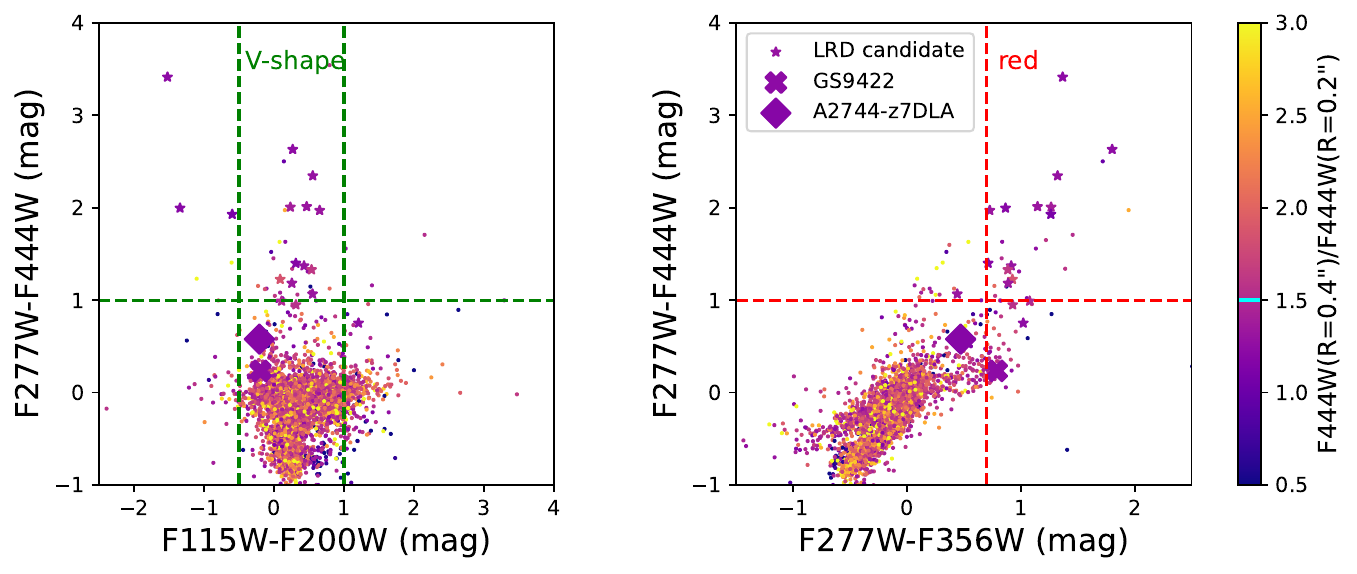}
 \vspace*{-1em}
 \caption{\small
High-redshift red galaxies from the UNCOVER program. The smallest points represent sources from the UNCOVER catalog \citep{uncover}, selected with $\rm F444W < 27.7\,mag$ and $\rm S/N > 14$ in the F444W band. Star symbols denote Little Red Dot (LRD) candidates from \citet{LRD_discover2}, cross symbol represent GS9422 from \citet{GS9422_3} while the diamond indicates the position of \galname, the galaxy presented in this work.
\textbf{Left:} NIRCam color-color diagram of F277W$-$F444W versus F115W$-$F200W, used to identify galaxies with blue rest-frame UV continua and red rest-frame optical continua—characteristic of a ``V-shaped'' spectral energy distribution (SED). The vertical dashed lines mark $\rm F115W - F200W = -0.5$ and $\rm F115W - F200W = 1.0$, and the horizontal line indicates $\rm F277W - F444W = 1.0$.
\textbf{Right:} NIRCam color-color diagram of F277W$-$F444W versus F277W$-$F356W, designed to select galaxies with red optical continua by applying color cuts between adjacent filters. The vertical line marks $\rm F277W - F356W = -0.7$, and the horizontal line again indicates $\rm F277W - F444W = 1.0$.
In both panels, all points are color-coded by the flux ratio $\rm F444W(R = 0.4'') / F444W(R = 0.2'')$, which is used to identify compact sources. The horizontal line in the color bar denotes a flux ratio of 1.5. All selection criteria follow those defined in \cite{LRD_discover2}. Notably, \galname appears to lie between the reliable LRD candidates and the bulk of the catalog galaxies.
 \label{fig:color selection}}
\end{figure*}

In Figure \ref{fig:color selection}, we present the well-detected red galaxies from the UNCOVER program. The 17 targets that satisfy the $(red1|red2)$ and $compact$ criteria are highlighted with star symbols, while \galname is marked with a diamond. We also plot GS9422 with photometric data from \citet{GS9422_3} for comparison.

Moreover, as noted by \cite{LRD_discover2,brown_dwarf}, brown dwarf stars are a major source of contamination in such selections. To mitigate this issue, \cite{LRD_discover2} introduced updated NIRCam-only selection criteria. These criteria incorporate the F277W–F356W color to identify SEDs with genuinely red continuum slopes, rather than objects whose colors are dominated by spectral breaks or emission lines. Overall, an object is considered red and compact only if it simultaneously meets all of the following four criteria:

$$compact\ red:$$
$$\rm{(-0.5 < F115W - F200W < 1.0)} $$
$$\rm{(F277W - F444W > 1.0)} $$
$$\rm{(F277W - F356W > 0.7)} $$
$$compact =\rm{ f_{F444W}(0.4'') / f_{F444W}(0.2'')} < 1.5$$

We show these criteria in Figure~\ref{fig:color selection} using dashed lines. While \galname satisfies the compactness requirement, it does not meet the ``V-shape'' color selection. Nevertheless, \galname clearly lies between the robust LRD candidates and the bulk of the catalog galaxies, occupying a similar region in the color--color space as GS9422, which may suggest comparable physical conditions.

\subsubsection{Host galaxy}
Recent observations reveal that many LRDs exhibit diffuse emission surrounding their compact, point-like centers \citep{decompose1, decompose3}. Further studies suggest that this diffuse component may trace both the accretion of halo gas and its subsequent conversion into stars \citep{decompose2}. 

Using \galfits\ to decompose the AGN and host galaxy of our source, we compare it with previous studies of high-redshift AGN host galaxies. According to \cite{galfits}, luminous blue quasars ($L_{5100} \gtrsim 10^{45} \, \rm erg\, s^{-1}$) reside in bulge-dominated galaxies (S\'{e}rsic index $n \approx 5$), while fainter red quasars inhabit disk-like galaxies (with $n \approx 1$). Based on the SED results, \galname is classified as a fainter source, with a S\'{e}rsic index of $n = 1.99$, which is closer to the red quasars. Additionally, red quasars tend to have a higher black hole-to-stellar mass ratio, suggesting the growth of black holes is faster than the growth of their host galaxies. Our source shows $\log (M_{\rm BH}/M_{\rm stellar}) = -1.58^{+0.45}_{-0.34}$, also between these two values, indicating a potential transitional stage. Lastly, while lower luminosity red quasars show a broad range of UV slopes ($\beta \approx -2$ to 4), luminous blue quasars have a narrow range ($\beta \approx 1.4$). \galname's steep UV slope aligns with the characteristics of fainter quasars.

\cite{decompose1} examines the properties of eight high-redshift LRDs, finding a host galaxy in only one, while four show extended, off-centered emission, possibly from starlight. \cite{decompose2} also observed significant extended emission in many LRDs, some of which can be explained by stellar models, while others suggest nebular emission. \galname exhibits a more dominant stellar component and likely nebular emission. We propose that \galname\ may represent a more evolved stage compared to typical LRDs, in which the AGN contribution is less dominant while the host galaxy plays a larger role, although the underlying process of stellar mass assembly remains consistent.

\subsubsection{Black hole mass}
\begin{figure*}
    \centering
    \includegraphics[width=0.8\linewidth]{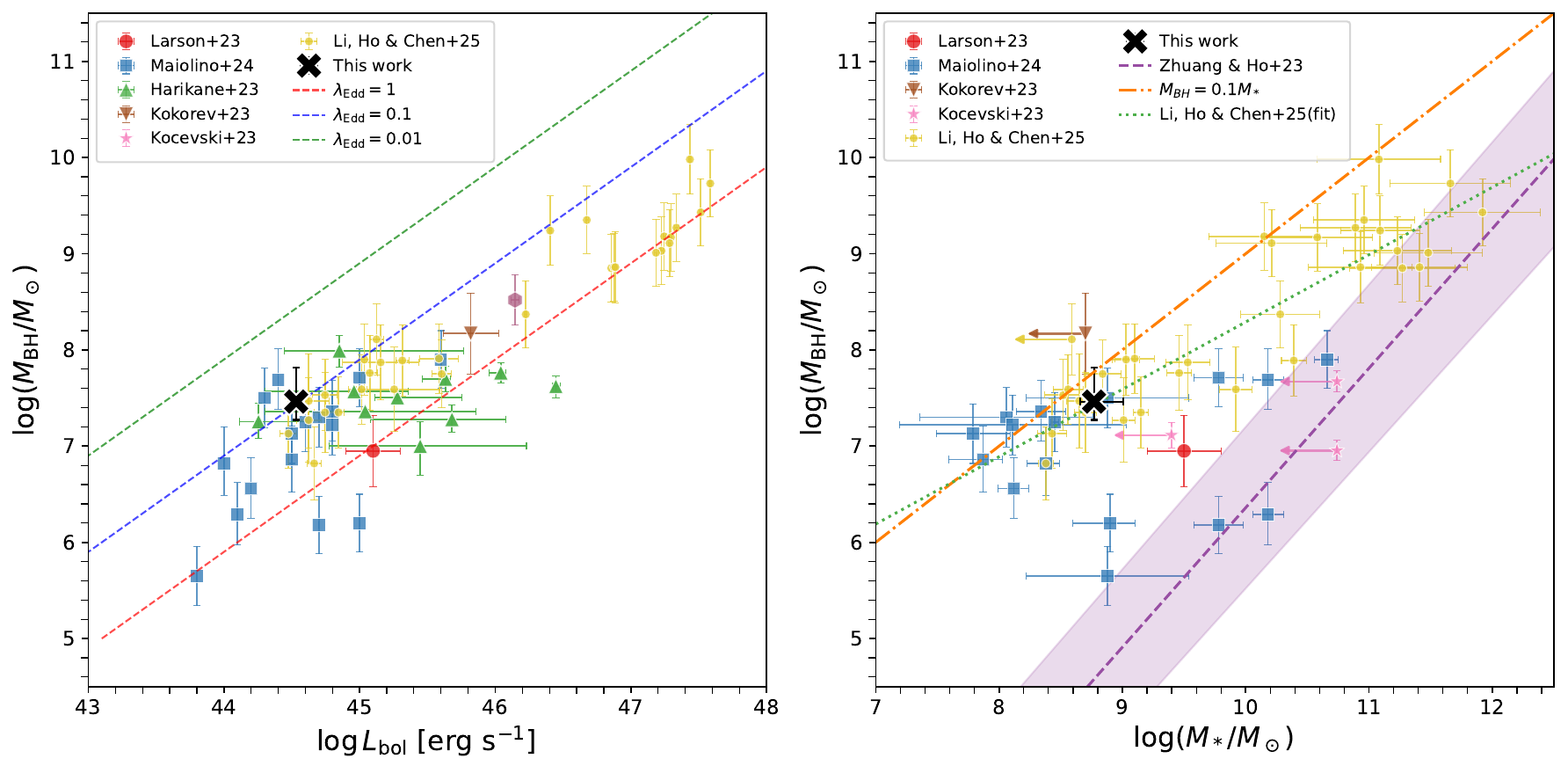}
   \caption{
\textbf{Left:} Relation between bolometric luminosity and black hole mass. We compare \galname\ with previously reported high-redshift AGNs and LRDs, including accreting supermassive black holes from \citet{BM3, maiolino+24}, faint high-redshift AGNs from \citet{harikane+23,    kocevski+23}, a massive black hole in a low-metallicity AGN from \citet{ubler+23}, a lensed galaxy with its quiescent companion from \citet{kokorev+23}, and a comprehensive AGN sample from several JWST surveys from \citet{galfits}. Our measurement is shown as a black cross. The orange, blue, and green dashed lines indicate Eddington ratios of 1, 0.1, and 0.01, respectively. Note that the bolometric luminosity of \galname\ is estimated from \bagpipes SED fitting; therefore, its uncertainty is not displayed.
\textbf{Right:} Black hole-to-stellar mass relation. The markers correspond to the same samples as in the left panel. The purple dashed line represents the best-fit relation that evolves to $z=0$ from \citet{evolutionary_host}, while the green dashed line corresponds to the linear fit result from \citet{galfits}. The orange dashed line marks the relation $M_{\rm BH} = 0.1\,M_\ast$. Note that the we use the result from \bagpipes as the stellar mass of \galname. The results indicate that \galname\ hosts a rapidly growing massive black hole, consistent with other comparison samples.
}
    \label{fig:Mbh}
\end{figure*}

We further considered additional physical properties commonly used to characterize AGNs. In the left panel of Figure \ref{fig:Mbh}, we present the estimated black hole mass, $\log (M_{\mathrm{BH}}/M_\odot)$, and bolometric luminosity, $\log L_{\mathrm{bol}}$, for our object, and compare them with other high-redshift galaxies and recently discovered LRDs. The black hole mass is derived from Equation \ref{BH_eq}, while the bolometric luminosity and the stellar mass are obtained from \bagpipes SED fitting. Dashed lines indicate Eddington ratios of 1, 0.1, and 0.01. The Eddington ratio exhibited by \galname\ falls between 0.01 and 0.1, consistent with most previous studies.  

In the right panel of Figure~\ref{fig:Mbh}, we show $\log (M_{\mathrm{BH}}/M_\odot)$ versus $\log (M_*/M_\odot)$ for our object. \galname\ lies close to the black hole mass–stellar mass relation for $z=4$--7 AGNs from \citet{galfits} (green dashed line): \begin{equation} \log\left(\frac{M_{\mathrm{BH}}}{M_\odot}\right) = 1.29 + 0.70 \log\left(\frac{M_\star}{M_\odot}\right) \end{equation} 
The local black hole mass–stellar mass relation, as estimated by \citet{evolutionary_host}, is also shown. It has been known that newly discovered high-redshift AGN tend to be overmassive relative to the stellar mass of their host galaxies, which cannot be explained entirely by the selection effect \citep[e.g.,][]{harikane+23,ubler+23,furtak+24,BlackTHUNDER,juodvzbalis+24}. Recent JWST observations suggest that such black holes are overmassive by a factor of $\sim$10–100 compared to local galaxies of similar stellar mass, and the inferred high-z BH-to-stellar mass relation is significantly different from that in the local universe \citep{pacucci+23}. For \galname\ at $z=6.87$, the results indicate the presence of a rapidly growing massive black hole, similar to other comparison samples.

\subsection{Massive accreting black hole} \label{SMBH}
According to the \galfits\ results, \galname\ contains an AGN component, with an inferred black hole mass of $\sim 2.9 \times 10^{7},M_\odot$. Given the presence of a DLA system dominating the nearest absorption, dense gas is likely located in the vicinity of, or within, the galaxy.

Widely accepted models of the ``Little Red Dots" suggest that cool gas contributes to the formation of the ``V-shaped" SED. \cite{lin_model} proposed that gas with temperatures of $T \sim 5000$–$6000$K enveloping the central SMBH produces thermalized emission. \cite{dense_gas} hold that a Balmer break can arise in AGN spectra if the accretion disk is deeply shrouded in dense neutral gas clumps. In these regions, collisions excite hydrogen atoms to the n=2 state, enabling efficient absorption of the AGN's continuum emission at wavelengths just below the Balmer limit. The UV band is likely dominated by either an AGN continuum or a young stellar population, and is sometimes absorbed by DLA systems \citep{BlackTHUNDER}. \galname does not exhibit a Balmer break in the optical band, indicating an absence of thermalized emission from the warm neutral medium. However, the presence of cool gas induces curvature in the UV continuum.

\cite{coevolution} proposed a hypothesis on how feedback influences star formation in star-forming galaxies hosting an AGN. Whether the feedback is positive or negative (i.e., whether cooling is effective or ineffective) depends on the column density of cooled gas behind shocks, because the column density within a galaxy regulates the relative timescales of dynamical processes and cooling, thereby influencing the effectiveness of SMBH feedback on their host galaxies. This model implies that the black hole grows and the AGN activity rises first, subsequently triggering star formation and ultimately suppressing it. \galname exhibits dominant stellar emission, along with a broad H$\alpha$ component indicating the presence of an AGN. It also shows a moderate star formation rate, consistent with the transitional phase predicted by this model.

In summary, the most plausible interpretation for our source is that the broad component of the H$\alpha$ emission line is produced by a supermassive black hole (SMBH) accreting dense gas in its immediate surroundings, while the host galaxy has experienced significant star-forming activity. This scenario is consistent with the models proposed by \citet{dense_gas} and \citet{coevolution}.

\section{Conclusion}\label{sec:conclu}
We report the discovery of a peculiar galaxy in the epoch of cosmic reionization, \galname, located in the A2744 field. We analyzed the morphology of this object using both \galfit\ and \galfits, finding that it exhibits a very compact structure. 
The morphological decomposition analysis indicates that \galname contains a central PSF component. When compared with the SED fitting results from \bagpipes and \cigale, this component is likely dominated by stellar emission, particularly in the rest-frame UV bands, whereas any AGN contribution becomes more significant only at longer wavelengths.

Its UV continuum shows Damped Lyman-alpha (DLA) absorption features, along with moderate Ly$\alpha$ emission, suggesting the presence of high-density neutral hydrogen (HI) along the line of sight.

A2744-z7DLA exhibits a relatively red color and compact morphology, but lacks the typical ``V-shaped" spectral profile, and thus would not be classified as the ``Little Red Dot" (LRD) candidate. In color space, it is situated between the majority of reionization-era galaxies and LRDs. The galaxy also exhibits a very high electron temperature and low metallicity, indicating an early evolutionary stage with limited dust enrichment, although it may represent a more evolved phase compared to typical LRDs.

The H$\alpha$ emission line of A2744-z7DLA exhibits a broad component, and the galaxy falls within the AGN region on the OHNO diagnostic diagram, suggesting it may be a Type I AGN. We estimate the FWHM of the broad component to be approximately $2721\ \mathrm{km\ s^{-1}}$, corresponding to a black hole mass of $M_{\rm BH} = 2.90^{+2.35}_{-1.28} \times 10^7\ M\odot$. We show the relations between bolometric luminosity and black hole mass, as well as the black hole-to-stellar mass relation, finding that A2744-z7DLA is consistent with other high-redshift AGNs and LRDs. These results support the scenario in which \galname\ is in an active black hole mass assembly phase. Its central supermassive black hole is likely accreting high-density neutral gas, which gives rise to the observed broadening of the emission lines.

Future observations will be essential to assess the AGN interpretation for this source. High–resolution spectroscopy would enable a more definitive search for broad-line components or high-ionization emission features such as C IV or He II. Such observations would provide critical evidence to either support or rule out the presence of an AGN in this compact system.

\begin{acknowledgments}
We thank the anonymous referee for very constructive comments that help improve the quality of this paper.
This work is supported by the National Key R\&D Program of China No.2025YFF0510603, the National Natural Science Foundation of China (grant 12373009), the CAS Project for Young Scientists in Basic Research Grant No. YSBR-062, the China Manned Space Program with grant no. CMS-CSST-2025-A06, and the Fundamental Research Funds for the Central Universities. XW acknowledges the support by the Xiaomi Young Talents Program, and the work carried out, in part, at the Swinburne University of Technology, sponsored by the ACAMAR visiting fellowship.
LCH was supported by the National Science Foundation of China (12233001) and the China Manned Space Program (CMS-CSST-2025-A09).
The authors sincerely thank the UNCOVER team (PIs: Labb\'e \& Bezanson; PID: GO-2561) for developing their observing program with a non-propriety period.
\end{acknowledgments}

\section*{Data Availability}
Some of the data presented in this article were obtained from the Mikulski Archive for Space Telescopes (MAST) at the Space Telescope Science Institute. The specific observations analyzed in this study can be accessed via \dataset[doi:	
10.17909/9chn-9q11]{https://doi.org/10.17909/9chn-9q11}. The data collection DOI was generated using the MAST DOI service\footnote{\href{https://archive.stsci.edu/publishing/doi}{https://archive.stsci.edu/publishing/doi}}.

\software{\galfits \citep{galfits}, 
\bagpipes \citep{bagpipes1,bagpipes2},
\lmfit \citep{lmfit},
\cigale \citep{cigale},
\msaexp \citep{msaexp},
\grizli \citep{grizli},
\psfex \citep{PSFEx},
\sextractor \citep{sextractor}}

\bibliographystyle{aasjournal}
\bibliography{reference}

\end{document}